\documentclass[aps,prl,notitlepage,longbibliography,twocolumn,superscriptaddress,10pt]{revtex4-2}
\usepackage[colorlinks=true,linkcolor=blue,citecolor=blue,urlcolor=blue]{hyperref}
\usepackage[german, english]{babel}
\usepackage[dvipsnames]{xcolor}
\usepackage{amsmath,amsfonts, amssymb, amsthm, dsfont}
\usepackage{bm}
\usepackage{graphicx}
\usepackage{tikz}
\usepackage{xspace}
\usepackage[export]{adjustbox}

\usepackage{enumerate}
\usepackage{subcaption}
\usepackage[justification=raggedright,format=plain]{caption}
\usepackage{bbold}
\usepackage{wasysym}
\usepackage{physics}

\newcommand{\calZ}{\mathcal{Z}}

\newcommand{\calN}{\mathcal{N}}

\newcommand{\mZ}{\mathcal{Z}}
\newcommand{\mNA}{\mathcal{N}_{A}}
\newcommand{\bA}{\overline{A}}

\newcommand{\sigmabar}{\overline{\sigma}}
\newcommand{\Abar}{{\bar{A}}}
\newcommand{\mR}{\mathcal{R}}

\begin{document}

\title{Persistent Topological Negativity in a High-Temperature Mixed-State}

\author{Yonna Kim}
\affiliation{Department of Physics, University of California, Santa Barbara, CA 93106, USA}	

\author{Ali Lavasani}
\affiliation{Kavli Institute for Theoretical Physics, University of California, Santa Barbara, CA, 93106}

\author{Sagar Vijay}
\affiliation{Department of Physics, University of California, Santa Barbara, CA 93106, USA}	

\begin{abstract}

We study the entanglement structure of the Greenberger–Horne–Zeilinger (GHZ) state as it thermalizes under a strongly-symmetric quantum channel describing the Metropolis-Hastings dynamics for the $d$-dimensional classical Ising model at inverse temperature $\beta$. This channel outputs the classical Gibbs state when acting on a product state in the computational basis.  When applying this channel to a GHZ state in spatial dimension $d>1$, the resulting mixed-state changes character at the Ising phase transition temperature from being long-range entangled to short-range-entangled as temperature increases. Nevertheless, we show that the topological entanglement negativity of a large region is insensitive to this transition and takes the same value as that of the pure GHZ state at any finite temperature $\beta>0$. We establish this result by devising a local operations and classical communication (LOCC) ``decoder" that provides matching lower and upper bounds on the negativity in the thermodynamic limit which may be of independent interest. This perspective connects the negativity to an error-correction problem on the $(d-1)$-dimensional bipartitioning surface and explains the persistent negativity in certain correlated noise models found in previous studies. Numerical results confirm our analysis. 
\end{abstract}

 \maketitle

{\bf Introduction:} The nature and fate of quantum entanglement in mixed-states of quantum many-body systems -- arising in thermal quantum matter or after a pure state experiences local decoherence -- have been recently investigated in order to understand the universal quantum correlations that can persist in open quantum systems \cite{wu2020entanglement,lu2020structure,lu2023characterizing,fan2024diagnostics,chen2023symmetry,chen2024separability,chen2024unconventional,sang2023mixed,lu2023mixed,lessa2024mixed,zhang2024quantum,lee2023quantum,de2022symmetry,zhu2023nishimori,chen2024unconventional,lee2022symmetry,wang2023topologically,wang2024anomaly,lee2022decoding}. Patterns of long-range-entanglement (LRE) in mixed quantum matter are intricately related to quantum many-body phases which can act as quantum error-correcting codes \cite{bao2023mixed,lu2020detecting,wang2023intrinsic}. Thermal states of quantum many-body systems are also now known to be efficiently classically-preparable at sufficiently high temperatures \cite{bakshi2024high}.

The entanglement negativity provides a well-studied diagnostic of bipartite entanglement in a mixed-state, which vanishes if the state can be prepared from a product state by local operations and classical communication (LOCC) across the  bipartition \cite{vidal2002computable}.  While the negativity of a region grows with the area of its boundary in a generic Gibbs state \cite{sherman2016nonzero}, contributions to the negativity which are independent to coarse-grained details of the bipartition signal the presence of LRE and can probe the presence of mixed-state topological quantum order \cite{lu2020detecting}.  Diagnostics of non-local mixed-state entanglement, such as the topological entanglement negativity \cite{lu2020detecting}, are challenging to study analytically, and the subtle behavior of mixed-state LRE under a finite-depth local quantum channel (FDLC) remains to be fully understood \cite{sang2023mixed,sang2024stability}.  

In this work, we investigate the fate of mixed-state quantum entanglement under a Metropolis-Hastings \cite{metropolis1953equation, hastings1970monte} channel describing the equilibration of a thermal \emph{classical} system, specifically that of a $d$-dimensional Ising model at inverse temperature $\beta$. When acting on a pure state in the computational basis, this channel prepares a classical thermal state of the $d$-dimensional Ising model. 

When the initial state is chosen to be a Greenberger–Horne–Zeilinger (GHZ) state, we show that the negativity of a subregion of the resulting mixed-state $\rho_{\beta}$ at any finite temperature $\beta > 0$ is asymptotically exactly that of the pure GHZ state in the limit that the boundary of the sub-region becomes large. In spatial dimensions $d>1$, the negativity is thus \emph{a constant, independent of geometric details of the bipartitioning surface and insensitive to the thermal phase transition}, as summarized in Eq. (\ref{eq:result_negativity}). Nevertheless, the entanglement properties of the mixed-state do change across this transition: we show that $\rho_{\beta}$ cannot be expressed as a convex sum of short-range-entangled (SRE) pure states in the ordered phase but is preparable using an FDLC acting on a product initial state in the disordered phase.  This observed behavior is in stark contrast to the negativity of the Gibbs state of a local, quantum many-body Hamiltonian which necessarily vanishes above a finite temperature \cite{bakshi2024high}.  

This result is established using the fact that the entanglement negativity is an entanglement monotone \cite{vidal2002computable,plenio2005logarithmic}, so that the negativity of a subregion $A$ of any state $\rho$, denoted $\mNA(\rho)$, cannot increase under an LOCC operation $\mR$ across the bipartitioning surface,
\begin{align}\label{eq:monotone}
    \mNA(\mR[\rho])\le \mNA(\rho).
\end{align}
We explicitly construct an LOCC operation that recovers the GHZ state at any finite temperature and for any bipartition in the asymptotic limit that the boundary of $A$ becomes large.  We show that for a contiguous bipartition, this channel is the decoder for a $(d-1)$-dimensional repetition code, which the two parties in the LOCC protocol must use to determine the precise unitary circuit that will recover the GHZ state. The well-known success of the $(d-1)$-dimensional repetition code up to a maximal error strength in dimensions $d > 1$ is intimately related to the recovery of the GHZ state, and thus the constant entanglement negativity, at any finite temperature. 

We comment on broader utility of this perspective for understanding mixed-state negativity by demonstrating that this can provide a simple understanding of why the topological entanglement negativity can remain a constant in correlated noise models (e.g. in \cite{wang2023intrinsic}).

{\bf Setup:} Consider $N$ qubits on a $d$-dimensional hypercubic lattice, in a state described by the density matrix
\begin{align}
{\rho}_{\beta} \equiv \frac{1}{\mZ_{\beta}}\sum_{\sigma}e^{-\beta H}\ket{\psi_{\sigma}}\bra{\psi_{\sigma}}. 
\label{eq:rho_beta}
\end{align}
Here, 
\begin{align}\label{eq:H_Ising}
H \equiv -J\sum_{\langle i,j\rangle}Z_{i}Z_{j}    
\end{align}
is the energy of a classical Ising model on the $d$-dimensional hypercubic lattice with ferromagnetic ($J>0$) nearest-neighbor interactions, while $\mZ_{\beta} \equiv \Tr(e^{-\beta H})$ is the corresponding partition function at inverse temperature $\beta$.  Furthermore, the state
$\ket{\psi_{\sigma}} \equiv (\ket{\sigma} + U\ket{\sigma})/\sqrt{2}$, 
where $\ket{\sigma} \equiv \ket{\sigma_{1},\ldots,\sigma_{N}}$ is a product state in the Pauli $Z$ basis, and $U\equiv \prod_{j}X_{j}$ is a unitary transformation which generates the Ising symmetry of Eq. (\ref{eq:H_Ising}).

The mixed-state $\rho_{\beta}$ is naturally obtained by starting with a GHZ state $\ket{\psi_{+}} = (\ket{\uparrow\cdots} + \ket{\downarrow\cdots})/\sqrt{2}$, which is an equal-amplitude superposition of the macroscopically-distinct all-up and all-down qubit configurations, and then applying a channel $\Phi_{\beta}$ which describes a Metropolis-Hastings algorithm \cite{metropolis1953equation,hastings1970monte} in which the energy $H$ is measured, and then a unitary operation $X_{j}$ is applied on a random qubit with probability $\min\{1, \exp(-\beta\,\Delta E_{j})\}$, where $\Delta E_{j}\equiv E_{\mathrm{new}} - E_{\mathrm{old}}$ is the energy difference between the new and old qubit configurations. 

Repeated measurements and feedback, starting from any product state in the Pauli $Z$ basis, will eventually produce a Gibbs ensemble of the $d$-dimensional classical Ising model at inverse temperature $\beta$ \cite{metropolis1953equation,hastings1970monte}.   We specifically define $\Phi_{\beta}$ to be  repeated application of the Metropolis-Hastings algorithm so as to produce a thermal steady-state when acting on the all-up state $\rho_{0}\equiv \ket{\uparrow\cdots}\bra{\uparrow\cdots}$, i.e.
\begin{align}
\Phi_{\beta}[\rho_{0}] = \frac{1}{\mZ_{\beta}}\sum_{\sigma}e^{-\beta H}\ket{\sigma}\bra{\sigma}.
\end{align}
Each operation that comprises the channel $\Phi_{\beta}$ manifestly commutes with the Ising symmetry transformation $U$.  Therefore, $U$ is a strong symmetry of this channel \cite{albert2014symmetries,lieu2020symmetry}, i.e.    $U\Phi_{\beta}[\rho]=\Phi_{\beta}[U\rho]$ $\forall \rho$, so that
\begin{align}
\Phi_{\beta}[\ket{\psi_{+}}\bra{\psi_{+}}] = \frac{1+U}{\sqrt{2}}\Phi_{\beta}[\rho_{0}]\frac{1+U}{\sqrt{2}} = \rho_{\beta}.
\end{align}

{\bf Mixed-State Entanglement:}
The state $\rho_{\beta}$ witnesses a thermal phase transition with increasing temperature in spatial dimensions $d>1$ between a ferromagnetically-ordered and a disordered state. At any temperature, the state $\rho_{\beta}$ has a strong Ising symmetry $U\rho_{\beta} = \rho_{\beta}$.  However, the ferromagnetically-ordered phase is also characterized by long-range-order in local operators which are charged under this symmetry $\Tr(\rho_{\beta}Z_{i}Z_{j}) \overset{|i-j|\rightarrow\infty}{\ne}0$.  These conditions are sufficient to establish  \cite{lu2023mixed} that this mixed-state cannot be described as an ensemble of SRE pure-states.  In Ref.\cite{chen2023symmetry} it has been argued that $\rho_\beta$ in the disordered phase can be written as an ensemble of SRE pure states. We prove a potentially stronger result  in the Supplemental Material \cite{Supp_Mat}, by showing that in the disordered phase, $\rho_\beta$  can be prepared by a FDLC acting on a product state and is thus SRE based on the even more restrictive definition of SRE mixed-states in Refs.\cite{ma2023average,ma2023topological,anshu2020circuit}.

\begin{figure}
    \includegraphics[width=0.4\linewidth]{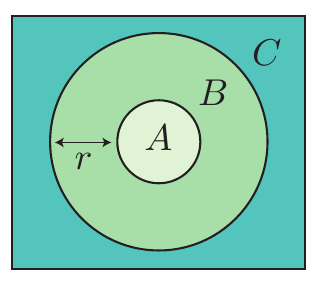}
    \caption{The $ABC$ partitioning used for computing the conditional mutual information $I(A:C|B)$.}
    \label{fig:ABC}
\end{figure}
{\bf Conditional Mutual Information:}
We now investigate entanglement properties of $\rho_{\beta}$ and their relation to the Ising phase transition. The conditional mutual information (CMI) is a measure of tripartite correlations in a state and has been used to study various aspects of entanglement in quantum systems \cite{levin2006detecting,kato2016information,ding2016conditional,fawzi2015quantum,lee2024universal,zhang2024nonlocal}, including probing mixed state quantum phases of matter \cite{sang2024stability}. 
Given subsystems $A$, $B$, and $C$, the CMI is defined as
\begin{align}
    I(A:C|B)
    &=S(AB)+S(BC)-S(B)-S(ABC).\label{eq:CMI}
\end{align}
Consider the partitioning shown in Fig. \ref{fig:ABC} with $r>0$, where $C$ is taken to be the complement of $AB$. Note that for any proper subset $A$ of qubits, $\rho_{\beta,A}=\tr_{\bar{A}}(\rho_\beta)=\rho_{\beta,A}^\text{Cl}$, where $\rho_\beta^\text{Cl}={e^{-\beta H}}/{\calZ_\beta}$ is the classical Gibbs state \footnote{This in particular shows that for mutual information we have $I(A:C)=I^\text{Cl}(A:C)$ provided that $r>0$, where $I^\text{Cl}$ is computed with respect to $\rho_\beta^\text{Cl}$.}. This shows that the first three terms in Eq. \eqref{eq:CMI} take the same value when computed for $\rho_\beta$ as for $\rho_\beta^\text{Cl}$. Moreover, it is straightforward to use  Eq.\eqref{eq:rho_beta} to show $S(\rho_\beta)=S(\rho_\beta^\text{Cl})-\log(2)$. Therefore $I(A:C|B)=\log(2)+I^\text{Cl}(A:C|B)$,
where $I^\text{Cl}(A:C|B)$ is the CMI computed for $\rho_\beta^\text{Cl}$. On the other hand, since $B$ is assumed to be non-empty, there is no interaction term in $H$ with support on both $A$ and $C$ and hence $I^\text{Cl}(A:C|B)=0$ \cite{clifford1971markov}, and
\begin{align}
    I(A:C|B)=\log(2),
\end{align}
independent of $r$, temperature $\beta$, and the system size. It is worth mentioning that in contrast, for the thermal Gibbs state of local quantum Hamiltonians, CMI decays exponentially with $r$ \cite{kato2019quantum,kuwahara2020clustering,kuwahara2024clustering}.

{\bf Entanglement Negativity:}
We now investigate the entanglement negativity of a subregion $A$ of the mixed-state (\ref{eq:rho_beta}), $\mNA(\rho_{\beta}) \equiv (||\rho_{\beta}^{T_{A}}||_{1} - 1)/2$.  We will show that for a region $A$, 
\begin{align}\label{eq:result_negativity}
\mNA(\rho_{\beta}) = \left\{\begin{array}{cc}
\displaystyle\frac{1}{2} - O(\exp(-|\partial A|))&\hspace{.1in}\beta>0\\
&\\
0&\hspace{.1in}\beta=0
\end{array}\right.
\end{align}
where $|\partial A|$ is the number of qubits on the boundary of $A$. The second result is easily established. At infinite temperature, $\rho_{\beta = 0} = (1 + U)/2^{N}$, so that $\rho_{\beta}^{T_{A}} = \rho_{\beta}$ and $\mNA(\rho_{\beta=0})=0$. 

The first result is determined as follows. The negativity is a convex function \cite{vidal2002computable,plenio2005logarithmic} so that 
\begin{align}\label{eq:convexity}
    \mNA(\rho_{\beta}) \le \frac{1}{\mZ_{\beta}}\sum_{\sigma}e^{\beta J\sum_{\langle i,j\rangle}\sigma_{i}\sigma_{j}} \mNA(|\psi_{\sigma}\rangle\langle\psi_{\sigma}|) = \frac{1}{2}
\end{align}
for any region $A$, since $|\psi_{\sigma}\rangle$ is related to the GHZ state $\ket{\psi_{+}}$ by single qubit gates, so that $\mNA(|\psi_{\sigma}\rangle\langle\psi_{\sigma}|) = \mNA(|\psi_{+}\rangle\langle\psi_{+}|)$.  The latter is exactly $1/2$ for any region $A$. The negativity is also an entanglement monotone \cite{vidal2002computable, plenio2005logarithmic} (see Eq. (\ref{eq:monotone})).  Here, we show that given the state $\rho_{\beta}$ for any $\beta > 0$, there is an LOCC operation $\mR$ with respect to any region $A$ and its complement $\bar{A}$ that recovers the GHZ state with a fidelity that approaches unity exponentially quickly in the size of the boundary of $A$.  Combining this with (\ref{eq:convexity}), allows us to establish Eq. (\ref{eq:result_negativity}).  

\begin{figure}
     \includegraphics[width=\columnwidth]{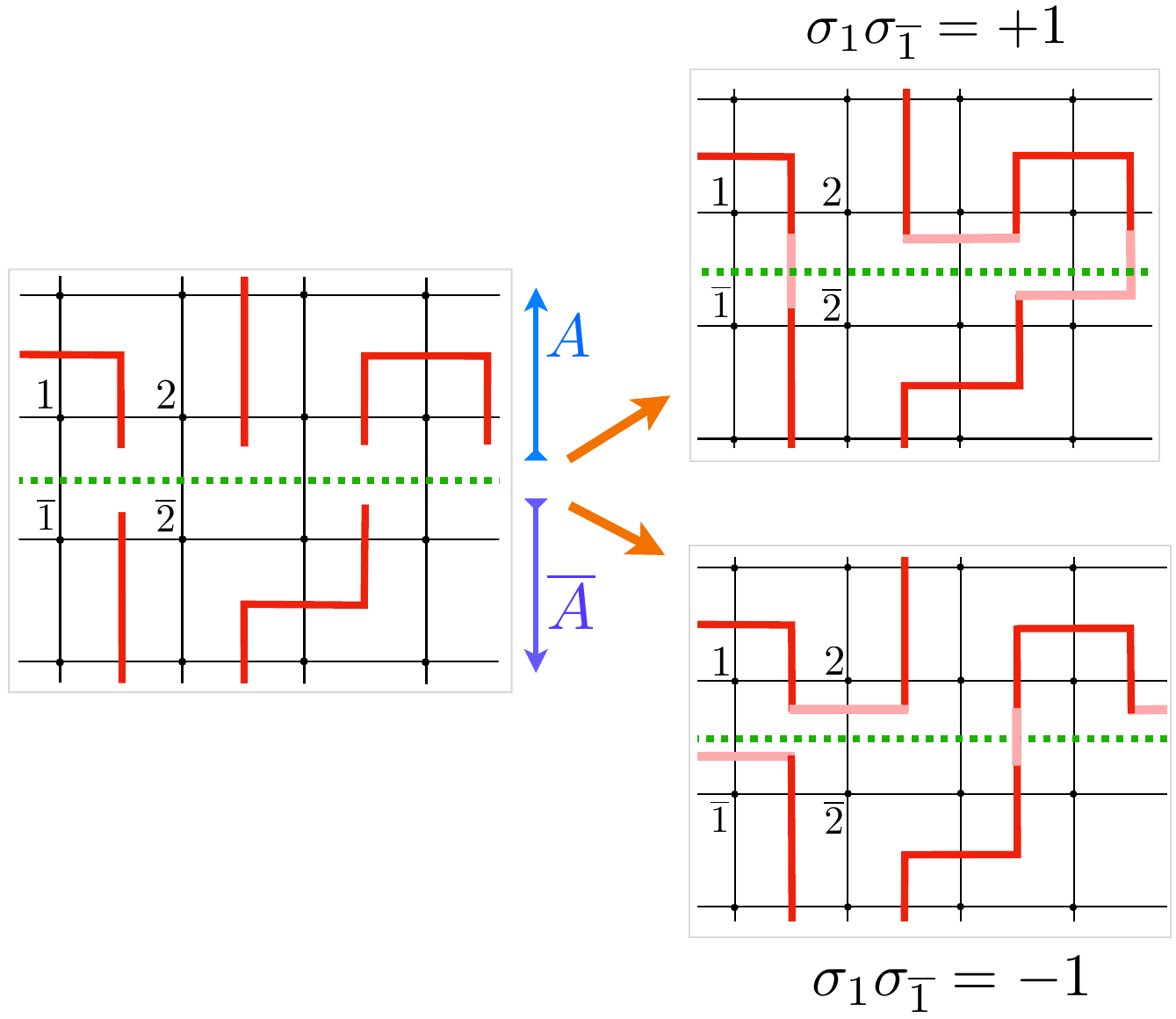}
     \caption{Measurements of domain walls that lie exclusively within $A$ or $\overline{A}$ lead to configurations such as the one shown in the left panel, with domain walls indicated by red lines on the dual lattice. There are two possibilities for how these domain wall connect across this interface, corresponding to the choices $\sigma_{1}\sigma_{\overline{1}} = \pm 1$.}
     \label{fig:LOCC}
\end{figure}

We construct the LOCC channel $\mR$ for a contiguous bipartition before generalizing this construction for a non-contiguous subregion $A$.  The state $\rho_{\beta}$ describes the ensemble of pure-states $\ket{\psi_{\sigma}}$ drawn with probability $e^{\beta J\sum_{\langle i,j\rangle}\sigma_{i}\sigma_{j}}/\mZ_{\beta}$. Given a state from this ensemble, we measure the operators $Z_{i}Z_{j}$ along all nearest-neighbor bonds lying exclusively within $A$ and $\bA$.  The resulting pattern of domain walls (see e.g. Fig. \ref{fig:LOCC}) -- is consistent with having drawn the state  (i) $\ket{\psi_{\sigma}}$ or the state (ii) $\prod_{j\in A}X_{j}\ket{\psi_{\sigma}}$ and leaves one classical bit's worth of ambiguity in determining the precise state.  

The preparation of the GHZ state thus maps onto a decoding problem in a classical $(d-1)$-dimensional repetition code living on the interface between $A$ and $\overline{A}$.  The classical bit of information is the relative alignment of any two qubits across the bipartition in the Pauli $Z$ basis (e.g. $\tau_{1}\equiv\sigma_{1}\sigma_{\overline{1}}$ in Fig. \ref{fig:LOCC}).  The measured ``syndromes" of the repetition code (e.g. $\tau_{1}\tau_{2} = (\sigma_{1}\sigma_{\overline{1}})(\sigma_{2}\sigma_{\overline{2}})$) and the Boltzmann weights for the $d$-dimensional Ising model are used to perform maximum-likelihood decoding to guess the correct state.  Let $w_{1}$ and $w_2$ denote the Boltzmann weights corresponding to states ($i$) and ($ii$)  respectively. We guess that the state we have is ($i$) with probability $p \equiv w_{1}/(w_{1}+w_{2})$ and ($ii$) with probability $1-p$.  We subsequently apply the appropriate single-site unitary gates to turn this into the GHZ state.
The LOCC channel is thus 
\begin{enumerate}
\item Measure $Z_{i}Z_{j}$ along all nearest-neighbor bonds lying exclusively within $A$ or $\bA$.
\item Let $\ket{\sigma}$ be a product state in the Pauli $Z$ basis which is consistent with the measurement outcomes. Apply the unitary $U_{1} = \prod_{j}X_{j}^{(1-\sigma_{j})/2}$ with probability $p$, or the unitary $U_{2} = \prod_{j\in A}X_{j}\prod_{j}X_{j}^{(1-\sigma_{j})/2}$ with probability $1-p$. 
\end{enumerate}

At any finite temperature, the relative probability $p/(1-p) = w_{1}/w_{2}$ scales exponentially in the size of the boundary of $A$, due to the fact that the energy of an Ising domain wall is extensive in its length.  Thus, the decoder for the repetition code should be successful at any finite temperature and fail exactly at infinite temperature when $w_{1} = w_{2} = 1/2$.  This is consistent with the known behavior of the repetition code with single-qubit bit-flip errors with probability $q$, for which perfect decoding is possible for any $q\le 1/2$. 

We may generalize this channel for the case where $A$ is a non-contiguous region.  Step 1 (above) is replaced by measurements of $Z_{i}Z_{j}$ across \emph{all} pairs of sites entirely within $A$ and entirely within $\bar{A}$.  As before, only one classical bit's worth of ambiguity remains after these measurements.  The quantum circuit that takes this state onto the GHZ state is then determined by a decoder for a repetition code with $|\partial A|$ sites.  Thus, decoding the correct quantum circuit should again be successful at any $\beta > 0$ as $|\partial A|$ becomes large, regardless of the geometry of the bipartition.  

For any bipartition, we show that the fidelity of the recovered state $R(\rho_{\beta})$ with the GHZ state is \cite{Supp_Mat}
\begin{align}\label{eq:locc_fidelity}
\mathcal{F}\left(\mR(\rho_{\beta}),\ket{\psi_{+}}\bra{\psi_{+}}\right) = \frac{1}{2}\left[1 - \Big\langle \tanh(\beta H_{\partial})\Big\rangle\right]
\end{align}
where $H_{\partial}$ denotes terms in the Ising Hamiltonian acting exclusively along bonds crossing the bipartition between $A$ and $\overline{A}$ and the expectation value is taken in the thermal state for the classical Ising model.  At any finite temperature, the energy $H_{\partial}$ is negative, and extensively large in the boundary of $A$, so that $\Big\langle \tanh(\beta H_{\partial})\Big\rangle \rightarrow 1$ as $A$ becomes large.

\begin{figure}
 \begin{subfigure}{0.4\textwidth}
     \includegraphics[width=\textwidth]{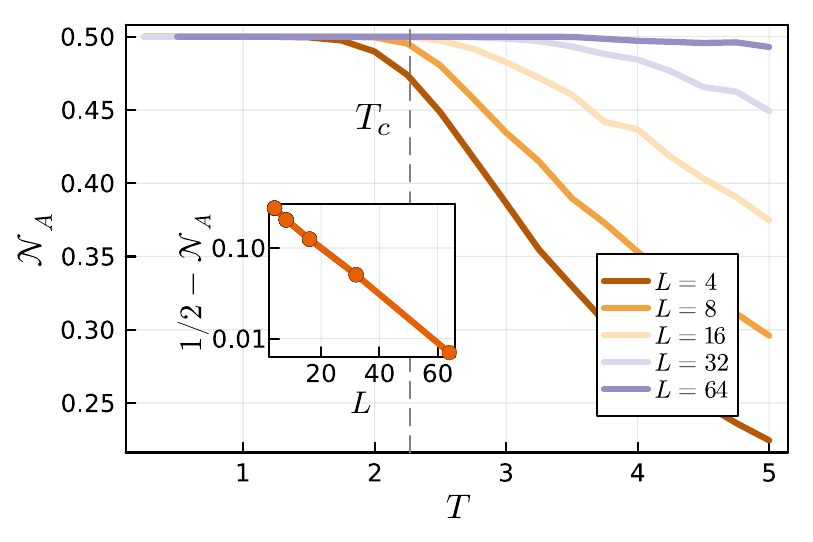}
     \caption{}
     \label{fig:N_half}
 \end{subfigure}
 \begin{subfigure}{0.4\textwidth}
     \includegraphics[width=\textwidth]{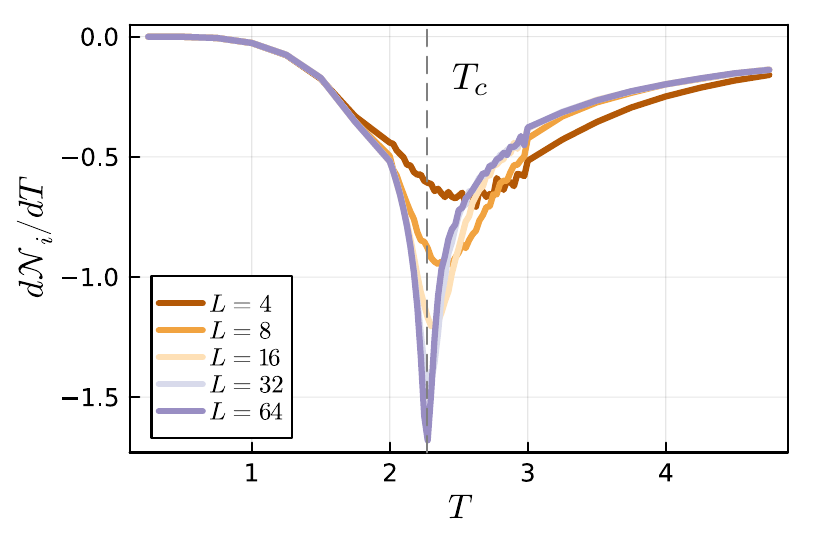}
     \caption{}
     \label{fig:dNdT_single}
 \end{subfigure}

 \caption{a) Entanglement negativity $\mathcal{N}_A$ versus temperature in the state $\rho_{\beta}$ in 
$d=2$, where subregion $A$ is taken to be half the system. The inset shows $1/2-\calN_A$ at $T=5$ versus $L$ in a semi-log plot, which shows that $\calN_A$ saturates exponentially quickly in the size of the boundary. 
 b) Derivative of entanglement negativity of a single site with respect to temperature, showing a singularity at the Ising phase transition temperature. 
}
\end{figure}

{\bf Numerical Results:} 
To numerically confirm this result, we obtain an alternate expression for the entanglement negativity of any bi-partition

\begin{align}
    \mathcal{N}_A[\rho_\beta]=\frac{1}{2}\expval{|\tanh(\beta H_\partial[\sigma])|}_q\label{eq:importance_sampling}
\end{align}
where the final expectation value is taken with respect to a distribution $q[\sigma]$, which is derived from the Boltzmann distribution for the Ising model, and is given in the Supplemental Material \cite{Supp_Mat}. We note that a similar result has been found before for a closely related set of mixed states \cite{tarunTalk}.

We present the numerical results of classical Monte Carlo simulations for a 2D system of qubits on an $L\times L$ square lattice with periodic boundary conditions. 
Fig.~\ref{fig:N_half} shows the negativity $\mathcal{N}_{A}(\rho_{\beta})$ when $A$ is taken to be the $L\times \frac{L}{2}$ cylinder. As is clear from Fig.~\ref{fig:N_half}, in the thermodynamic limit, the negativity is equal to $1/2$ even above the Ising critical temperature (marked by the dashed line in Fig.\ref{fig:N_half}), as expected from the expression in \eqref{eq:importance_sampling}. The fact that the negativity of subregions with extensive boundary is $1/2$ irrespective of temperature, means that if one defines the topological negativity as the constant contribution to the negativity as in Ref. \cite{lu2020structure}, one would find that the resulting mixed state has $\mathcal{N}_\text{topo}=-1/2$ even above the critical temperature where $\rho_\beta$ is an SRE mixed-state. \cite{werner1989quantum,hastings2011topological,ma2023average,chen2024separability,chen2023symmetry} (cf. Ref.~\cite{lu2020detecting}). 

Although the thermal phase transition at $T_c$ has no effect on the negativity of subregions with extensive boundaries, the negativity of finite subregions (while continuous) becomes singular at $T=T_c$. Fig.~\ref{fig:dNdT_single} shows the derivative of the negativity of a single site with respect to temperature which shows a singularity at $T=T_c$, which is consistent with the singularity of the heat capacity at the $d=2$ classical Ising phase transition.

\begin{figure}
     \includegraphics[width=.52\columnwidth]{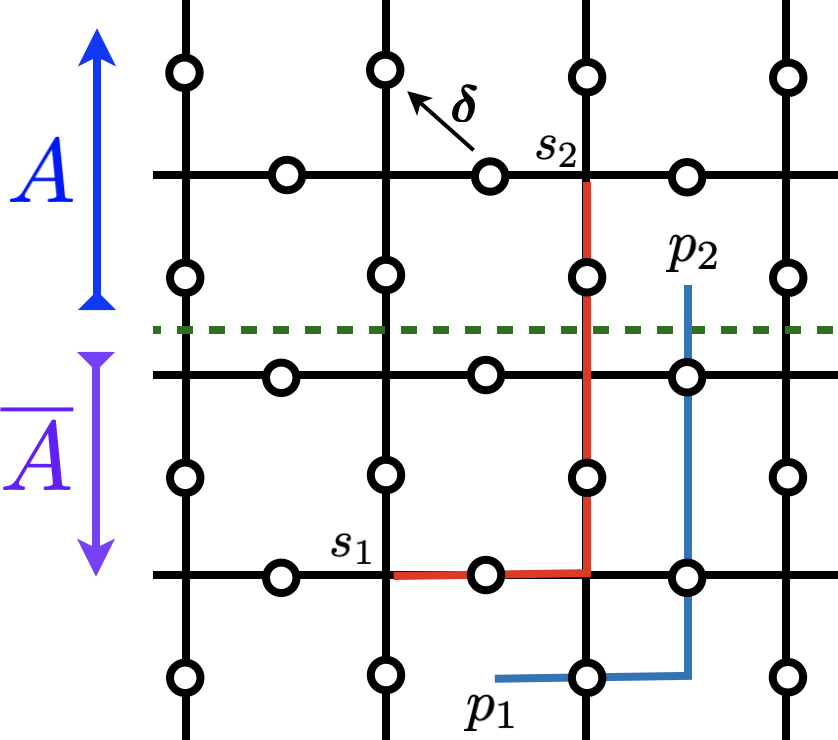}
     \caption{Dephasing of the toric code by the local operators $Z_{j+\boldsymbol{\delta}}X_{j}$ at each lattice site $j$ as studied in Ref. \cite{wang2023intrinsic}. Correlations in the syndromes permit perfect preparation of a mixed-state within the ground-space of the toric code using LOCC across $A$ and $\overline{A}$.}
     \label{fig:ZX_Dephasing}
\end{figure}

{\bf Discussion and Outlook:} Using LOCC to recover states with a known pattern of LRE may provide another perspective that can be used to investigate mixed-state LRE under decoherence. For correlated decoherence, this is apparent in specific examples.  The two-dimensional toric code dephased by local operators $Z_{j+\boldsymbol{\delta}}X_{j}$ at each site $j$, with the vector $\boldsymbol{\delta}$ given in Fig. \ref{fig:ZX_Dephasing}, was studied in Ref. \cite{wang2023intrinsic}.  Quantum state trajectories under this channel only contain patterns of $\mathbb{Z}_{2}$ charge and flux which are separated by $\boldsymbol{\delta}$.  Therefore, for any dephasing strength, it is possible to use LOCC across the bipartition in Fig. \ref{fig:ZX_Dephasing} to deterministically recover the toric code ground-state. The $\mathbb{Z}_{2}$ charge and flux lying on the bipartitioning surface, which cannot be directly measured in the LOCC protocol, can be perfectly inferred from stabilizer measurements in $A$ and $\overline{A}$ due to the noise correlations.  The negativity for this bipartition and for any dephasing strength \cite{wang2023intrinsic} is thus \emph{exactly} that of the toric-code ground-state \cite{castelnovo2013negativity}. Another example for which the LOCC perspective presented here could be insightful is the study of entanglement structure in the toric code with single-site $Y$ decoherence \cite{tuckett2019tailoring,ellison2024towards}. In this case, while the noise model itself is not correlated, the LOCC protocol can benefit from the correlations that are present in the syndromes due to the extensively many strong symmetries of this channel \cite{tuckett2019tailoring,hauser2024information}. The detailed study of this model and other correlated noise models is left to a future work. 

We identify other directions implied by our results.  First, it is interesting to consider the fate of mixed-state entanglement under thermalization by more generic quantum channels which are strongly-symmetric.
More concretely, we may consider thermalizing the GHZ state via a strongly symmetric implementation of the \emph{quantum} Metropolis-Hastings dynamics \cite{temme2011quantum,yung2012quantum} for the $d$-dimensional transverse field quantum Ising model. The resulting mixed state would be the projection of the quantum Gibbs state onto the symmetric sector. A similar LOCC protocol to recover the GHZ state can be used to relate the entanglement negativity to the performance of a $(d-1)$-dimensional repetition code under a \emph{coherent} correlated noise model. Energetic considerations can be used to argue about the success of the LOCC protocol in turning the resulting mixed state into the GHZ state. This lower bounds the negativity of the resulting mixed-state and suggests that the negativity is non-vanishing at high temperatures, though this remains to be confirmed. 

Second, while the persistence of topological negativity at high temperatures in our work can be understood as a consequence of the repetition code possessing maximum error threshold, it is interesting to search for settings where a similar line of argument would lead to a different code with a finite threshold. Such a system would exhibit a finite temperature phase transition in negativity, which may not necessarily coincide with the standard order-disorder thermal phase transition. 

Lastly, given that the topological negativity is unable to detect the phase transition in the complexity of the mixed-state $\rho_\beta$ (from LRE to SRE), it is interesting to look for other quantities that would be able to capture the complexity of the quantum phase transition in our system as well as in more general settings. A possible candidate would be the convex-roof extension of the conditional mutual information as introduced in Ref. \cite{wang2024analog} using the same $ABC$ partitioning used above. 

{\bf Acknowledgments:}
We thank Yimu Bao, Matthew Fisher, Tarun Grover, Tibor Rakovszky, Subhayan Sahu, Shengqi Sang and Yaodong Li for useful discussions. This research was supported in part by grant NSF PHY-2309135 to the Kavli Institute for Theoretical Physics (KITP). Use was made of computational facilities purchased with funds from the National Science Foundation (CNS-1725797) and administered by the Center for Scientific Computing (CSC). The CSC is supported by the California NanoSystems Institute and the Materials Research Science and Engineering Center (MRSEC; NSF DMR 2308708) at UC Santa Barbara. SV is partly supported as an Alfred P. Sloan Research Fellow.


\bibliography{refs}

\onecolumngrid
\newpage
\appendix\label{Supp_Mat}
\begin{center}
    {\Large Supplemental Material}
\end{center}

\section{Entanglement of $\rho_{\beta}$}\label{app:SRE_rho_beta}
Here, we show that the state (Eq. \ref{eq:rho_beta}) is short-range-entangled (SRE) throughout the disordered phase at sufficiently high temperatures, and in any spatial dimension.  First, observe that this state may be written as
\begin{align}\label{eq:rho_b_app}
\rho_{\beta} = \sum_{\mu}p_{\mu}\ket{\phi_{\mu}}\bra{\phi_{\mu}}
\end{align}
where $\ket{\phi_{\mu}}\equiv \rho_{\beta}^{1/2}\ket{\mu}/ \sqrt{p_{\mu}}$ and $p_{\mu}\equiv \bra{\mu}\rho_{\beta}\ket{\mu}$, while $\ket{\mu}$ is a product state in the Pauli $X$ basis.  Because of the strong symmetry of this state, $\Tr(\rho_{\beta}U) = 1$, where $U \equiv\prod_{j}X_{j}$.  As a result, each pure state appearing in (\ref{eq:rho_b_app}) must satisfy $U\ket{\phi_{\mu}} = \ket{\phi_{\mu}}$.  

Furthermore, observe that
\begin{align}
\rho_{\beta}^{1/2} = \frac{1}{\sqrt{\mZ_{\beta}}}\sum_{\sigma}e^{-\beta H/2}\ket{\psi_{\sigma}}\bra{\psi_{\sigma}}.
\end{align}
It is evident from this expression that $[Z_{i}Z_{j},\rho_{\beta}^{1/2}]=0$ for any $i$, $j$.  As a result, any state $\ket{\phi_{\mu}}$ appearing in (\ref{eq:rho_b_app}) can be obtained by applying single-site unitary gates on the reference state $\ket{\phi_{+}} = \rho_{\beta}^{1/2}\ket{+\cdots +}/\sqrt{p_{+}}$, where $\ket{+\cdots +}$ is a product state in which $X_{j}=+1$ for each qubit:
\begin{align}
\ket{\phi_{\mu}} = \prod_{j}Z_{j}^{\frac{1-\mu_{j}}{2}}\ket{\phi_{+}}. 
\end{align}
We note that $p_{\mu} = \bra{\mu}\rho_{\beta}\ket{\mu} = \bra{+\cdots +}\prod_{j}Z_{j}^{\frac{1-\mu_{j}}{2}}\rho_{\beta}\prod_{j}Z_{j}^{\frac{1-\mu_{j}}{2}}\ket{+\cdots +} = \bra{+\cdots +}\rho_{\beta}\ket{+\cdots +} = p_{+}$.  As a result, $p_{\mu} = 1/2^{N-1}$ where $N$ is the total number of qubits.  To summarize, we may write (\ref{eq:rho_b_app}) as
\begin{align}
\rho_{\beta} &= \frac{1}{2^{N-1}}\sum_{\mu \,\,\mathrm{s.t.}\,\,\prod_{j}\mu_{j}= +1}\prod_{j}Z_{j}^{\frac{1-\mu_{j}}{2}}\ket{\phi_{+}}\bra{\phi_{+}}\prod_{j}Z_{j}^{\frac{1-\mu_{j}}{2}}.
\end{align}

We now re-cast the summation by introducing Ising degrees of freedom $\sigma_{ij}$ along bonds of the hypercubic lattice, and letting $\mu_{j} \equiv A_{j}[\sigma] \equiv \prod_{i\in NN(j)}\sigma_{ij}$, where the product is taken over sites that are nearest-neighbor to $j$.  With periodic boundary conditions, $\prod_{j}A_{j}[\sigma] = +1$, and as a result, the constraint on the summation $\prod_{j}\mu_{j} = +1$ is naturally satisfied by this replacement.  With this replacement, we may write
\begin{align}
\rho_{\beta} &= \frac{1}{2^{N_{b}}}\sum_{\sigma} \prod_{j}Z_{j}^{\frac{1-A_{j}[\sigma]}{2}}\ket{\phi_{+}}\bra{\phi_{+}}\prod_{j}Z_{j}^{\frac{1-A_{j}[\sigma]}{2}}.
\end{align}
Here, $N_{b} = d\cdot N$ is the total number of bonds on the $d$-dimensional hypercubic lattice.  The pre-factor ensures the normalization of the density matrix \footnote{Another way to derive this pre-factor is to observe that for a given configuration of $\mu_{j}$ satisfying the constraint $\prod_{j}\mu_{j} = 1$, there are $2^{N_{b}}/2^{N-1}$ different configurations of $\sigma_{ij}$ such that $\mu_{j} = A_{j}[\sigma]$.}. Using the fact that $Z_{j}^{\frac{1-A_{j}[\sigma]}{2}} = \prod_{i\in\mathrm{NN}(j)}Z_{j}^{\frac{1-\sigma_{ij}}{2}}$, we may write $\prod_{j}Z_{j}^{\frac{1-A_{j}[\sigma]}{2}} = \prod_{\langle i,j\rangle}(Z_{i}Z_{j})^{\frac{1-\sigma_{ij}}{2}}$, where the product in the second expression is taken over distinct, nearest-neighbor bonds.  As a result,
\begin{align}
\rho_{\beta} &= \frac{1}{2^{N_{b}}}\sum_{\sigma} \prod_{\langle i,j\rangle}(Z_{i}Z_{j})^{\frac{1-\sigma_{ij}}{2}}\ket{\phi_{+}}\bra{\phi_{+}}\prod_{\langle i,j\rangle}(Z_{i}Z_{j})^{\frac{1-\sigma_{ij}}{2}}.
\end{align}
We conclude that $\rho_{\beta}$ may be obtained by dephasing $Z_{i}Z_{j}$ on each bond of the lattice, with \emph{maximal} strength, a manifestly finite-depth, local quantum channel.

Finally, we may argue that the state $\ket{\phi_{+}}$ is SRE when the system is in a disordered phase.  We do this by first writing
\begin{align}\label{eq:phi_plus}
\ket{\phi_{+}} = \sqrt{2^{N-1}}\rho_{\beta}^{1/2}\ket{+\cdots +} = \sqrt{\frac{2^{N-1}}{\mZ_{\beta}}}\sum_{\sigma}e^{-\beta H/2}\ket{\psi_{\sigma}}\langle{\psi_{\sigma}|+\cdots +}\rangle = \frac{1}{\sqrt{\mZ_{\beta}}}\sum_{\sigma}e^{-\beta H/2}\ket{\sigma}.
\end{align}
We now observe that (\ref{eq:phi_plus}) is the ground-state of the Hamiltonian
\begin{align}
H = -\sum_{j}Q_{j}
\end{align}
where
\begin{align}
    Q_{j}\equiv -X_{j} + \prod_{i\in\mathrm{NN}(j)}e^{-\beta J Z_{i}Z_{j}}.
\end{align}

\section{LOCC Operations and Recovering the GHZ State}

Here we will show the existence of an LOCC protocol $\mR$  with respect to bi-partition $A$ and $\Abar$, which transforms state $\rho_\beta$ back to the GHZ state in the thermodynamic limit. 
The LOCC protocol consists of two steps:

\begin{enumerate}
    \item 
    Measure $Z_iZ_j$ on each link that is entirely in $A$ or in $\Abar$ subregions, finding the outcome $\mu_{ij}=\pm 1$. This does \textbf{not} include the links across the boundary, $\partial A$.

    \item For any given set of measurement outcomes $\mu=\{\mu_{i,j}\}$, there exist two possible states which have domain wall configuration consistent with $\mu$. 
     We pick one of the two states based on a coin toss weighted with their respective Boltzmann weight in $\rho_\beta$ (Eq.\eqref{eq:rho_beta}) and
      apply a product of Pauli $X_i$ operators to eliminate the domain walls in that assumed configuration.
\end{enumerate}
In the following, we will derive Eq.\eqref{eq:locc_fidelity} for the fidelity between the GHZ state, $\ket{\psi_+}$, and the output state $\mR(\rho_\beta)$.
\subsection{Fidelity}

Let $\{M_\mu\}$ be the projective measurements that measure whether there is a domain wall or not on the links which are entirely
in region $A$ or $\bar{A}$,

\begin{align}
    M_\mu \equiv \prod_{<i,j>\in A, \bar{A}} \frac{1+\mu_{i,j}Z_iZ_j}{2},
\end{align}
where $\mu_{ij}$ is the measurement outcome of measuring $Z_iZ_j$ and the product only includes nearest-neighbor sites which are both in $A$ or both in $\Abar$.
Note that,
\begin{align}\label{eq:M_psi}
    M_\mu \ket{\psi_\sigma} = \prod_{<i,j>\in A, \bar{A}}\delta_{\mu_{i,j},\sigma_i\sigma_j} \ket{\psi_\sigma}.
\end{align} 
\begin{figure}
    \centering
    \includegraphics[width=0.4\linewidth]{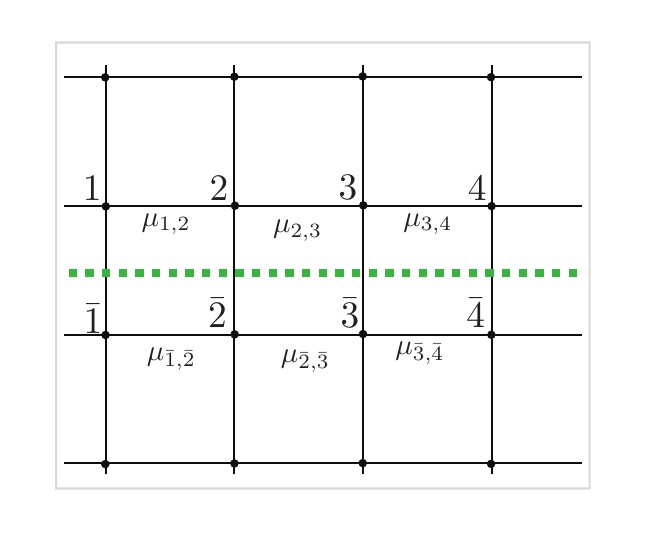}
    \caption{Labeling the boundary sites in $A$ and $\Abar$}
    \label{fig:lattice_notation}
\end{figure}
A set of measurement outcomes $\mu$ specifies a certain domain wall configuration in $A$ and $\Abar$, with some domain walls ending on the boundary $\partial A$. Let $1\in A$ and $\bar{1}\in\Abar$ be two adjacent sites across the boundary and define $\tau=\pm 1$ to be a marker for whether or not there is a domain wall across the edge connecting $1$ and $\bar{1}$, i.e. $\tau\equiv \sigma_1 \sigma_{\bar{1}}$ (see Fig.~\ref{fig:LOCC}). Note that $\mu$ and $\tau$ together uniquely specify the global domain wall configuration, because there should be even number of domain walls across each plaquette of the square lattice. In particular if we label the sites along the boundary as shown in Fig.\ref{fig:lattice_notation}, we have,
\begin{align}
    \sigma_i \sigma_{\,\bar{i}}=\tau\prod_{j=1}^{i-1}\mu_{j, j+1}\mu_{\bar{j}, \bar{j}+1}\label{eq:sigma_tau_mu}.
\end{align}
Let $X_{\mu,\tau}$ be the appropriate product of single qubit Pauli-$X$s that corresponds to the domain wall configuration specified by $\mu$ and $\tau$. In particular, if $M_\mu \ket{\psi_\sigma}=\ket{\psi_\sigma}$, we have
\begin{equation}\label{eq:X_mt_psi}
    X_{\mu,\tau}\ket{\psi_\sigma} \equiv 
    \begin{cases}
        \ket{\psi_+} & \text{if $\sigma_1\sigma_{\bar{1}}=\tau$}\\
        \prod_{j\in A}X_j\ket{\psi_+} & \text{if $\sigma_1\sigma_{\bar{1}}\neq\tau$}.
    \end{cases}
\end{equation}
Given the measurement outcomes $\mu$, the decoder $\mR$ applies $X_{\mu,\tau}$ with $\tau=\pm 1$ chosen randomly with probability $P_{\mu,\tau}$. The probability $P_{\mu,\tau}$ is chosen to be proportional to the Boltzmann weight of the spin configuration specified by $\mu$ and $\tau$ and according to Eq.\eqref{eq:sigma_tau_mu} can be written as,
\begin{align}
    P_{\mu,\tau} = \frac{e^{\beta J \tau \sum_{i\ge 1} \prod_{j=1}^{i-1} \mu_{j, j+1}\mu_{\bar{j}, \bar{j}+1}}}{\sum_{\tau'=\pm 1} e^{\beta J \tau' \sum_{i\ge 1} \prod_{j=1}^{i-1} \mu_{j, j+1}\mu_{\bar{j}, \bar{j}+1}}}.
\end{align}
Putting everything together, we can compute the average ``decoding" fidelity,

\begin{align}
    F = \mel{\psi_+}{\mR(\rho_\beta)}{\psi_+}= \sum_{\sigma,\mu,\tau} \frac{e^{-\beta H[\sigma]}}{\mathcal{Z}_\beta}P_{\mu,\tau}\abs{\bra{\psi_+}X_{\mu,\tau}M_\mu\ket{\psi_\sigma}}^2.
\end{align}
Using Eqs.\eqref{eq:M_psi} and \eqref{eq:X_mt_psi} this simplifies to
\begin{align}
    F = \sum_{\sigma,\mu,\tau}\frac{e^{-\beta H[\sigma]}}{\mathcal{Z}_\beta} P_{\mu,\tau} \prod_{<i,j>\in A, \bar{A}}\delta_{\mu_{i,j},\sigma_i\sigma_j} \delta_{\sigma_1\sigma_{\,\bar{1}},\tau}.
    \label{F}
\end{align}
Plugging the expression for $P_{\mu,\tau}$ into Eq. \eqref{F}, we get

\begin{align}
    F =& \sum_{\sigma,\tau}\frac{e^{-\beta H[\sigma]}}{\mathcal{Z}_\beta} \frac{e^{\beta J\tau\sum_{i\geq1}\sigma_1\sigma_{i}\sigma_{\,\bar{1}}\sigma_{\,\bar{i}}}}{\sum_{\tau'=\pm1} e^{\beta J\tau'\sum_{i\geq1}\sigma_1\sigma_{i}\sigma_{\,\bar{1}}\sigma_{\,\bar{i}}}} \delta_{\sigma_1\sigma_{\,\bar{1}},\tau}\\
     =& \sum_{\sigma} \frac{e^{-\beta H[\sigma]}}{\mathcal{Z}_\beta} \frac{e^{\beta J\sum_{i\geq1}\sigma_i\sigma_{\,\bar{i}}}}{\sum_{\tau'=\pm1} e^{\beta \tau'\sigma_1 \sigma_{\,\bar{1}} J\sum_{i\geq1}\sigma_i \sigma_{\,\bar{i}}}} \\
     =&\frac{1}{2} \sum_\sigma \frac{e^{-\beta H[\sigma]}}{\mathcal{Z}_\beta} \left(1-\tanh(\beta J\sum_{i\geq1}\sigma_i \sigma_{\,\bar{i}})\right)\\
     =& \frac{1}{2}\left[1-\Big\langle \tanh(\beta H_\partial)\Big\rangle\right]
\end{align}
where $H_\partial$ is the boundary Hamiltonian between $A$ and $\bar{A}$. 

\subsection{Mapping to the $d-1$ repetition code}

The problem of choosing which domain wall configuration to assume can be understood as a decoding problem for a $(d-1)$-dimensional classical repetition code. To see this, note that the decoding task in a classical error correcting code is essentially figuring out what error has happened based on a given set of syndromes. In particular the decoding task in the classical repetition code on $N$ bits can be described as follows. Consider an \textit{error vector} $e\in \{1,-1\}^N$ such that $e_i=-1$ if a bit flip has happened on the $i$'th bit and $e_i=1$ otherwise. 
The goal of the decoder is to figure out $e$, knowing only the syndrome values, i.e. $e_i e_j$ for adjacent $i$ and $j$ on the appropriate graph, e.g. the $d$-dimensional square lattice in the standard $d$-dimensional repetition code. Moreover, the decoder assumes an underlying noise model, i.e. a random distribution from which the vector $e$ is drawn. For example, the independent and identically distributed (iid) noise model assumes each component $e_i$ is $-1$ with probability $p$ and $1$ with probability $1-p$, independent of other components. 

In our setting finding the domain wall configuration is clearly the same task, with $e_i$ defined to be $e_i\equiv \sigma_i \sigma_{\,\bar{i}}$. Note that we cannot know the value of $e_i$, but we know the value of $e_i e_j=\mu_{i,j}\,\mu_{\bar{i},\bar{j}}$ for adjacent $i$ and $j$. At low temperatures, the domain walls are rare so most $e_i$s would be $1$, and as one increases the temperature more $e_i$s would be $-1$. The underlying noise model is given by $P_{\mu,\tau}$ as described in the previous section. This noise model is not iid, but rather a correlated noise model. However, as we argue in the next section, the repetition code is able to guess $e$ correctly with probability $1$ in the thermodynamic limit unless $\beta=0$, at which point each $e_i=\pm 1$ completely at random. 

\section{Threshold for the Repetition Code}
Here we argue that the repetition code decoding under the noise distribution $P_{\mu,\tau}$ is successful with probability $1$ for any $\beta>0$ in the thermodynamic limit. First we review the argument for maximum error correction threshold in the presence of the iid noise and then we argue that the same should remain true for more generic noise models. We use quantum notation for convenience although the problem we are considering is inherently classical. 

Consider an initial state $\ket{\Psi}$ undergoing some errors, leaving it in the mixed state
\begin{align}
    \rho = \sum_e p_e~e\ket{\Psi}\bra{\Psi}e 
\end{align}
where $e$ is a product of $X$ bit flips and $p_{e}$ is the probability of having that error. We rewrite the sum to distinguish between the errors that leave $\rho$ with fewer than $N/2$ bit flips and errors with higher weights:
\begin{align}
    \rho = \sum_{e: |e|<N/2}p_e~e\ket{\Psi}\bra{\Psi}e + \sum_{e:|e|\geq N/2}p_e~e\ket{\Psi}\bra{\Psi}e.
\end{align}
The error correction channel, $\mR$, measures the syndromes and between two possible error configuration consistent with a syndrome, applies the one with the smaller weight. Therefore it always successfully corrects the errors when there are less than $N/2$ bit flips. Hence,
\begin{align}
    \mR(\rho)=& \sum_{e: |e|<N/2} p_e~\mR(e\ket{\Psi}\bra{\Psi}e) + \sum_{e:|e|\geq N/2}p_e ~\mR(e\ket{\Psi}\bra{\Psi}e)\\
    =& \sum_{e: |e|<N/2} p_e\ket{\Psi}\bra{\Psi} + \sum_{e:|e|\geq N/2}p_e~\mR(e\ket{\Psi}\bra{\Psi}e).
\end{align}
The fidelity of this post-error correction state with the original state is lower bounded as,
\begin{align}
    F =& \bra{\Psi}\mR(\rho)\ket{\Psi}\\
    =& \sum_{e: |e|<N/2} p_e +  \sum_{e:|e|\geq N/2}p_e~\bra{\Psi}\mR(e\ket{\Psi}\bra{\Psi}e)\ket{\Psi} \geq \sum_{e: |e|<N/2} p_e\label{eq:lower_bnd}
\end{align}
where the lower bound is because the second term in the expression above is non-negative. 

Now we will show that this lower bound is equal to one in the thermodynamic limit. First, we rearrange the sum to be over $|e|=k$, the number of bit flips error, and because  $p_k$ is a binomial distribution,
$
    p_k = \binom{N}{k}p^k(1-p)^{N-p}.
$
In the limit of large $N$, we can approximate the binomial distribution as a Gaussian distribution with mean $Np$ and standard deviation $\sqrt{Np(1-p)}$ 
via the central limit theorem. Since $N/2$ is $\frac{N/2-Np}{\sqrt{Np(1-p)}}=O( \sqrt{N})$ standard deviations away from the peak of the Gaussian, we see that 
$
    \lim_{N\rightarrow\infty} \sum_{k=0}^{N/2-1} p_k = 1
$
when $p<1/2$. Since fidelity is upper-bounded by one, the fidelity is indeed equal to one in the thermodynamic limit. Thus, the repetition code will work successfully when $p<1/2$. 

For the more general case where $e$ is drawn based on the $P_{\mu,\tau}$ distribution, the bound in Eq.\eqref{eq:lower_bnd} is still valid. Therefore, to show perfect recovery we only need to show that with probability one, less than half of the edges contain a domain wall. Intuitively, this is true because for $\beta>0$, not having a domain wall on an edge is energetically favorable. To make it rigorous, one can define indicator variables $I_i=(1-e_i)/2$ in terms of which the number of domain walls can be written as $k=\sum_{i}I_i$. When $\beta >0$ it is more probable for $I_i$ to be $0$ rather than $1$ and hence the expectation value of $k/|\partial A|$ is less than $1/2$. Noting that $\lim_{|i-j|\to \infty}\text{Cov}(I_i,I_j)=0$, the weak law of large numbers for weakly dependent variables \cite{cacoullos2012exercises} shows that $k/|\partial A|$ concentrates around its mean, which is less than $1/2$ when $\beta>0$ in the thermodynamic limit. 

\section{Negativity Spectrum}\label{apx:negativity_spectrum}
Here we calculate the entanglement negativity between a sub-region $A$ and its complement $\bar{A}$, in the state $\rho_\beta$. 
Let $\sigmabar$ denote the spin configuration that is related to $\sigma$ by flipping every spin. Accordingly, the state $\ket{\psi_\sigma}$ can be written as $\ket{\psi_\sigma}=\frac{\ket{\sigma}+\ket{\sigmabar}}{\sqrt{\sqrt{2}}}$ and Eq.~\eqref{eq:rho_beta} can be rewritten as,

\begin{align}
    \rho_\beta = \frac{1}{2\mathcal{Z}_\beta}\sum_\sigma e^{-\beta H[\sigma]}(\ket{\sigma}\bra{\sigma} + \ket{\bar{\sigma}}\bra{\bar{\sigma}} +\ket{\sigma}\bra{\bar{\sigma}} +\ket{\bar{\sigma}}\bra{\sigma}),
\end{align}
where $H[\sigma]=-J \sum_{\expval{i,j}} \sigma_i \sigma_j$ is the classical Ising energy of the spin configuration $\sigma$. Taking the partial transpose of $\rho_\beta$ with respect to sub-region $A$ yields,
\begin{align}
    \rho_\beta^{T_A} &= \frac{1}{2\mathcal{Z}_\beta}\sum_\sigma\left[ e^{-\beta H[\sigma]}\left(\ket{\sigma}\bra{\sigma} + \ket{\bar{\sigma}}\bra{\bar{\sigma}} \right) +e^{-\beta H[\sigma_A,\sigma_{\Bar{A}}]} \left(\ket{\Bar{\sigma}_A,\sigma_{\Bar{A}}}\bra{\sigma_A, \bar{\sigma}_{\Bar{A}}} +\ket{\sigma_A, \bar{\sigma}_{\Bar {A}}}\bra{\Bar{\sigma}_A,\sigma_{\Bar{A}}}\right)\right]\nonumber\\
    &=\frac{1}{2\mathcal{Z}_\beta}\sum_\sigma\left[ e^{-\beta H[\sigma]}\left(\ket{\sigma}\bra{\sigma} + \ket{\bar{\sigma}}\bra{\bar{\sigma}} \right) +e^{-\beta H[\sigma_A,\sigmabar_{\Bar{A}}]} \left(\ket{\Bar{\sigma}}\bra{\sigma} +\ket{\sigma}\bra{\Bar{\sigma}}\right)\right]
\end{align}
where $\sigma_A$ and $\sigma_{\bar{A}}$ denote the spin configuration $\sigma$ restericted to regions $A$ and $\bar{A}$ respectively. Thus, the eigenvalues and eigenstates of $\rho_A^{T_A}$ are
\begin{align}
    \lambda_{\sigma;\pm}=\frac{e^{-\beta H[\sigma]}\pm e^{-\beta H[\sigma_A,\bar{\sigma}_{\Bar{A}}]}}{\mZ_{\beta}}, \hspace{.5in} \ket{\psi_\sigma^{(\pm)}}\equiv \frac{\ket{\sigma}\pm\ket{\Bar{\sigma}}}{\sqrt{2}}.
\end{align}

We can now calculate the entanglement negativity  from the sum of all the negative eigenvalues. Firstly,
it is only the $\lambda_{\sigma;-}$ eigenvalues which can be negative. 
We can also see that $\lambda_{\sigma;-}$=$\lambda_{\bar{\sigma};-}$ from the $\mathbb{Z}_2$ symmetry of the Hamiltonian and they correspond to the same eigenvector, up to some phase. 
Moreover, 
note that $\lambda_{\sigma;-}=-\lambda_{\sigma_A,\bar{\sigma}_{\bar{A}};-}$ so for a given spin configuration, either $\lambda_{\sigma;-}$ or $\lambda_{\sigma_A,\bar{\sigma}_{\bar{A}};-}$ would be negative. Thus, we account for the two incidences of double counting with an additional 1/4 and get
\begin{align}
    \mathcal{N}_{A}[\rho_\beta] = \frac{1}{4\mZ_{\beta}}\sum_\sigma\abs{e^{-\beta H[\sigma]} - e^{-\beta H[\sigma_A,\Bar{\sigma}_{\Bar{A}}]}}.
\end{align}
Factoring out $e^{-\beta H[\sigma]}$ we get
\begin{align}
    \mathcal{N}_{A}[\rho_\beta] &= \frac{1}{4\mZ_{\beta}}\sum_\sigma e^{-\beta H[\sigma]}\abs{1 - e^{-\beta( H[\sigma_A,\Bar{\sigma}_{\Bar{A}}]- H[\sigma])}}\nonumber\\
    &=\frac{1}{4}\left\langle\abs{1 - e^{2\beta H_\partial[\sigma])}}\right\rangle\label{eq:negativity_thermal_epxrSM}.
\end{align}
 One may now use classical Monte Carlo simulations to compute the thermal average in Eq.\eqref{eq:negativity_thermal_epxrSM}.
However, care must be taken with sampling the spin configurations, since the expression inside the absolute value could be exponentially large for configurations with an exponentially small Boltzmann weight, in such a way that the overall contribution to the expectation value is $O(1)$.

The aforementioned issue can be avoided by using a technique known as importance sampling.  Let $q[\sigma]$ denote the probability distribution defined as $q[\sigma_A,\sigma_\Abar]=\frac{1}{2}p[\sigma_A,\sigma_\Abar]+\frac{1}{2}p[\sigma_A,\sigmabar_\Abar]$, where $p[\sigma]=e^{\beta J\sum_{\langle i,j\rangle}\sigma_{i}\sigma_{j}}/\mathcal{Z}_{\beta}$. To sample from $q[\sigma]$, one can simply sample from $p[\sigma]$ and then with $1/2$ probability, flip the spins in $\Abar$. We may thus write Eq.\eqref{eq:negativity_thermal_epxrSM} as,
\begin{align}
    \mathcal{N}_A[\rho_\beta]
    &=\frac{1}{4}\sum_\sigma q[\sigma]\frac{p[\sigma]}{q[\sigma]}|1-e^{2\beta H_\partial[\sigma]}|\nonumber\\
    &=\frac{1}{2}\expval{|\tanh(\beta H_\partial[\sigma])|}_q.\label{eq:importance_sampling}
\end{align}
where the final expectation value is taken with respect to the distribution $q$. The advantage of Eq.\eqref{eq:importance_sampling} compared to Eq.\eqref{eq:negativity_thermal_epxrSM} is that the expression inside the expectation value is now bounded by unity everywhere, so that contributions from rare configurations are no longer a concern.

\end{document}